\documentclass[journal=ancac3,manuscript=article]{achemso}
\usepackage[version=3]{mhchem} 
\usepackage{float}
\usepackage{graphicx}
\usepackage{chemformula} 
\usepackage[T1]{fontenc} 

\author{Matan Revah}
\author{Andre Yaroshevsky}
\author{Yuri Gorodetski}
\email{yurig@ariel.ac.il}
\affiliation[Ariel University]
{Ariel University, Mechanical Engineering \& Mechatronics department and Electrical Engineering \& Electronics department, Ariel, 40700, Israel}

\title[An \textsf{achemso} demo]
  {Spin-locking metasurface for surface plasmon routing}


\begin{document}

\begin{tocentry}
\includegraphics[height=3.5cm, keepaspectratio] {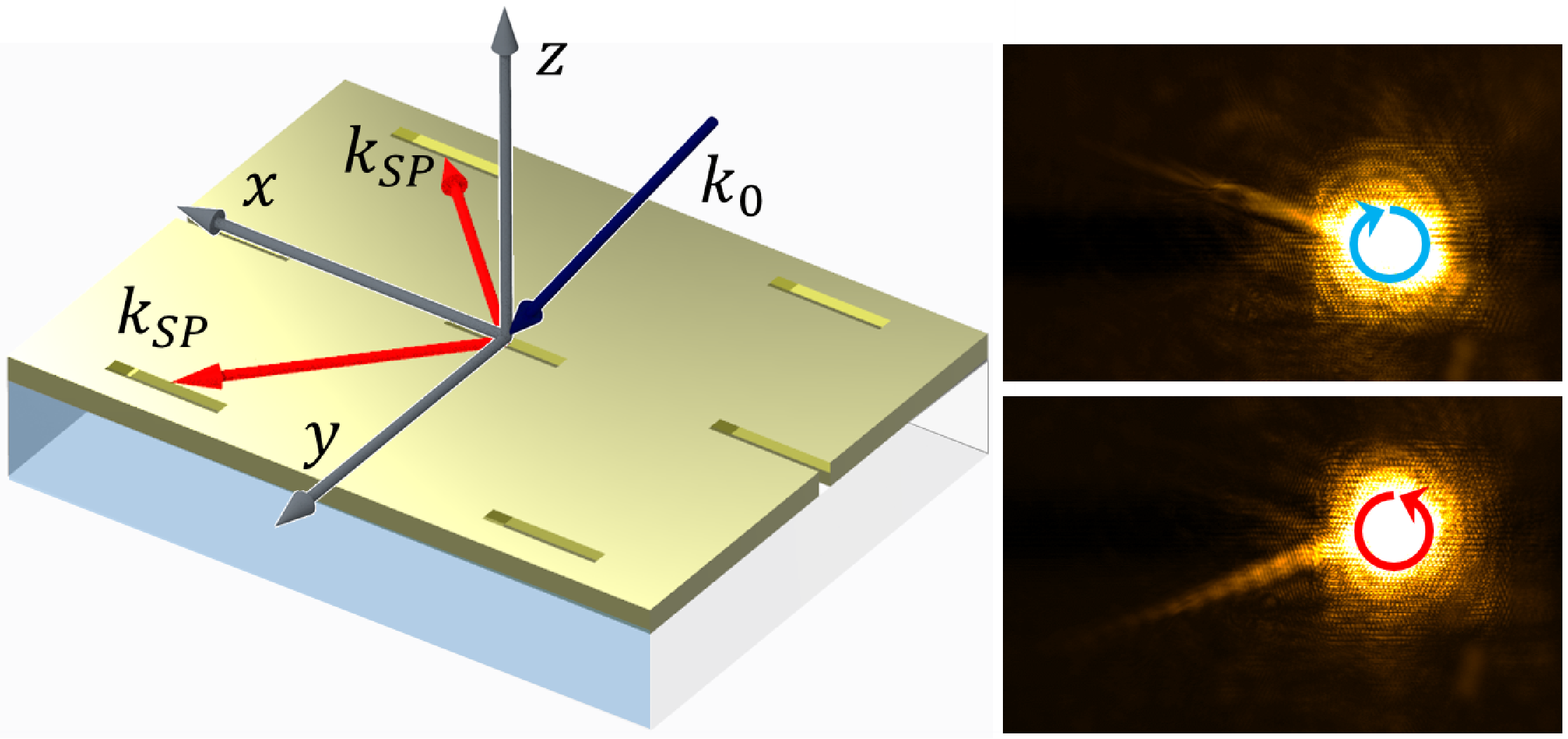}
	\end{tocentry}

\begin{abstract}
The spin-dependent routing of surface plasmons on metal by use of mirror symmetric periodic metasurface is presented and analyzed. 
We incorporate the intrinsic transverse spin angular momentum of the plasmonic wave in order to efficiently couple the incident light carrying circular polarization to a desired direction. 
The combination of the oblique incidence of polarized light with the accurately designed momentum matching properties of the grating provides a perfect way to achieve a spin-locking metasurface that routes the plasmonic beams.

\end{abstract}

\section{Introduction}

The increasing desire for nanophotonics integrated devices has led to an intensive investigation of surface waves such as surface plasmons (SP) due to their unique optical properties and compatibility with nanoelectronics \cite {ebbesen1998, Barnes2003,zayats2005nano}. 
While using metals as SP guiding material has numerous limitations there have been proposed various types of metamaterials - materials with effective properties - in order to achieve better signal-to-noise ratio and longer propagation range \cite {zheludev2012, cai2009optical, pendry2006, liu2011metamaterials}. 
In practice, this is achieved by nanostructuring a material in order to enhance local light-matter interactions in the nanoscale \cite {engheta2006}. 
Since the SP propagation is mostly sensitive to the properties of the metal-insulator interface a metasurface has been suggested as a plausible counterpart of metamaterials for some plasmonic applications \cite {yu2014flat, lin2014dielectric, litchinitser2012}. 
Different types of nanopatterned metallic surfaces have been proposed as functional metasurfaces capable of locally modulating the plasmonic wavefront phase and amplitude. 
Upon these, the metasurfaces providing an externally controlled near-field manipulation seem to be particularly appealing for nanophotonic devices \cite {papaioannou2016, kildishev2013planar}. 
Intrinsic polarization selectivity of the plasmonic wave makes it possible to use a plasmonic spin-orbit interaction for spin-based metasurfaces \cite {gorodetski2008, gorodetski2009, huang2012}. 
Such structures exhibit different behavior when excited by right-handed or left-handed circularly polarized light (RCP/LCP). 
In these systems the space-variant rotation of the structure unit-cell induces a geometric Berry phase of the plasmonic wavefront which results in a spin-dependent near-field distribution \cite {pancharatnam1956, berry1987, bomzon2002, bliokh2015}. 
This effect has been widely studied and various promising nanophotonic applications have been proposed \cite {huang2013, jiang2018}. 
Nevertheless, recently it was suggested that the SPs could carry a so-called transverse spin (TS) angular momentum which resulted from the relative quarter period phase lag between the longitudinal and the transverse field components \cite {Bliokh2014, aiello2015}. This TS was shown to be locked to the wave propagation direction and it had already been demonstrated that by illuminating a single slit in a metallic surface it was possible to create a projection of the longitudinal spin (LS, circular polarization handedness) onto the TS of the SP and excite unidirectional plasmonic wave \cite {OConnor2014, Lee2012Role,miroshnichenko2013}.
In this work we present a novel type of a spin-locking metasurface based on the LS-to-TS coupling and enhanced by the accurately designed momentum matching. 
Our structure collectively excites an SP wavefront in a desired direction depending on the incident circular state.
In contrast with other spin-based metasurfaces, here we demonstrate a periodic array of \textit{uniform apertures with full mirror symmetry}. 
The spin-locking is achieved through the incident beam inclination combined with apertures periodicity which allows a flexible design and relatively simple realization. 
We experimentally demonstrate the functionality of our device by using pulsed laser and measure the temporal dynamics of the plasmonic pulses excited by the metasurface.  

\section{Spin-locking metasurface}

When the SP wave propagates on a planar metal-air interface in $x$ direction its TS is given as 
\begin{equation}
  \textbf{s}_{\perp} \propto \frac{Re\textbf{k} \times Im \textbf{k}}{(Re\textbf{k})^2}
\label{TSdef}
\end{equation}
where $\textbf{k} = k_{SP}\mathbf{\hat{x}}+i\kappa \mathbf{\hat{z}}$ is the the complex valued evanescent wave vector with $\kappa = \sqrt{k_{SP}^2 - k_0^2}$, $k_0 = 2\pi/\lambda_0$ is the vacuum wavenumber and $k_{SP}$ is the in-plane plasmonic wavenumber 
\cite{Bliokh2015tr}.
As stated previously the transverse spin results from the rotation of the resultant of the vectorial plasmonic field, $\textbf{E}_{SP} = \textit{E}_{p}(\hat{\textbf{x}}-i\chi\hat{\textbf{z}})$ in a transverse plane and solely arises from the amplitude ratio between the longitudinal and the transverse field components $\chi$. 
Accordingly, $s_{\perp}$ is locked to the SPs' propagation direction and can appear with a single handedness. 
This has led to suggest a scheme for spin-dependent unidirectional plasmonic excitation \cite {miroshnichenko2013, OConnor2014} where the incidence geometry provides a considerable projection of the LS onto the TS of the plasmonic wavefront. 
Hereafter we refer to this effect as a ``longitudinal to transverse spin (LTS) coupling'' . 
One experimental way to achieve that was by using a single slit as a launching structure illuminated by an inclined Gaussian beam. 
Then, by choosing the inclination angle an optimal coupling conditions could be obtained. 
Nevertheless, such a system suffers from a spatially broadband behavior and requires a very precise localization of the illumination on the slit. 
Moreover, in the paraxial regime required for quazi-planewave excitation only a central part of the beam (the one falling directly on the slit) takes part in the launching process.

\begin{figure}[htbp]
	\centering
		\includegraphics[width=8.5cm, keepaspectratio] {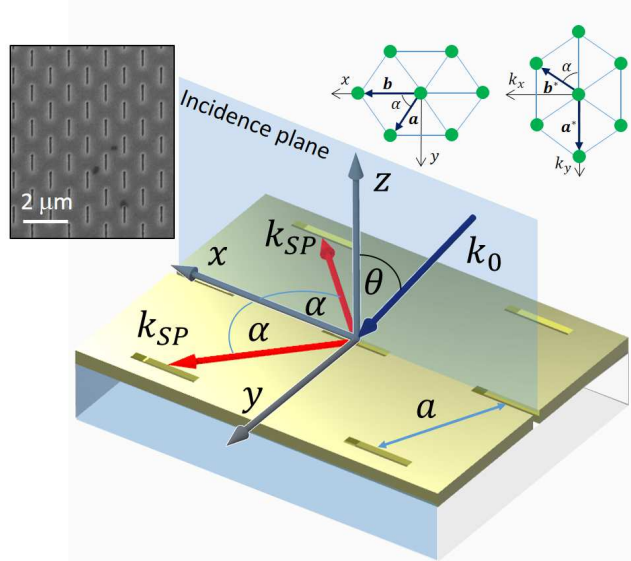}
		\label{Fig1}
		\caption{ The metasurface physical principle. The light is incident at an angle $\theta$ and couples to SPs propagating in the metasurface plane along the angle $\alpha$. The periodicity of the grating is $a$. The inset shows the direct and the reciprocal lattice corresponding to our metasurface.}
	\end{figure}

These issues make the system much less efficient and inappropriate for the practical use. 
Instead we suggest using a periodic structure of short and narrow ($100nm$ wide) nanoslits as a spin-locking metasurface. 
The system is designed to perform as follows. 
We bear in mind that an individual slit illuminated by a planewave at the incidence angle $\theta$ produces two plasmonic wings propagating at angle $\alpha$ with respect to the slit that can be derived via momentum matching condition as, $\cos \alpha = \frac{\lambda_{SP}}{\lambda_0} \sin \theta $. 
Instead of the single slit we now consider the array of slits placed roughly along the SP propagation direction at a distance providing the perfect phase matching. 

These requirements lead to a rhombic lattice of slits with a diagonal angle $\alpha$ (see Fig. 1). Previously some research has been performed in order to modify the resonant properties of SP excitation by use of these structures \cite{humphrey2014plasmonic, zhou2010toward}. In a periodic lattice, however, the momentum matching is different from the case of a single slit and should be derived by calculating the $k$-space of the array, i.e. its reciprocal lattice \cite{guo2017geometry}. 
Inset in Fig. 1 shows the geometry of our rhombic lattice represented by the pair of primitive vectors, 
$\textbf{a} = a\sin\alpha \hat{\textbf{y}}+a\cos\alpha \hat{\textbf{x}}$, 
$\textbf{b} = b \hat{\textbf{x}}$.
The corresponding primitive vectors in the $k$-space are then 
$\textbf{a}^{\ast} = \frac{2 \pi}{a\sin \alpha} \hat{\textbf{k}}_y $ and 
$\textbf{b}^{\ast} = \frac{2 \pi}{2 a}\left (\frac{\hat{\textbf{k}}_x}{\cos\alpha}-\frac{\hat{\textbf{k}}_y}{\sin\alpha}\right)$.    
We use these vectors to achieve the momentum matching condition for oblique incidence at angle $\theta$:  
\begin{equation}
  \left|\textbf{b}^{\ast}\pm\frac{2 \pi}{\lambda_0}\sin{\theta}\hat{\textbf{x}}\right| = \frac{2 \pi}{\lambda_{SP}} 
\label{Period}
\end{equation}

We note that the $\pm$ sign choice depends on the desired tilting of the beam with respect to the plasmonic propagation direction. 
This point will be discussed later in the paper. Clearly, the higher is the illumination tilting angle the larger momentum mismatch should be compensated which requires larger initial grating period. 
However, the tilt in our system is limited by the NA of the objective  (we used NA = 0.45). 
 Therefore we have studied nine structures with the diagonal angles of $\alpha = 67^{\circ}/45^{\circ}/30^{\circ}$ each existing with three periods, $a = 700nm/900nm/1200nm$. 
The table below shows the calculated illumination angle for each one of the lattice geometries according to the momenum matching condition in Eq. 2.

\begin{table}[h!]
\centering
\begin{tabular}{ |p{2cm}|p{2cm}|p{2cm}|p{2cm}|  }

 \hline
                & $67^{\circ}$            &  $45^{\circ}$          & $30^{\circ}$           \\
 \hline
 $700nm$        & $\theta = -25^{\circ}$  & $\theta = -7^{\circ}$  & $\theta = -22^{\circ}$  \\
 $900nm$        & $\theta = -9.6^{\circ}$ & $\theta = 12^{\circ}$  & $\theta = 3^{\circ}$\\
 $1200nm$       & $\theta = 7.5^{\circ}$  & $\theta = 27^{\circ}$  & $\theta = 24.5^{\circ}$  \\ 
 \hline
\end{tabular}
\label{table:1}
\caption{Incidence angles for different lattice geometries}
\end{table}

Those parameters were used for our grating design in order to generate different types of spin-locking metasurfaces. 
Note that the negative angles correspond to the case when the tilt of the illumination is in the opposite direction to the SP propagation.
In the following section we present our experimental observation of collective LTS coupling by spin-locking metasurfaces and show the dependence of this effect upon various geometric parameters of the structure. 

\section{Experimental results}

The metasurfaces were fabricated using focused ion beam (FIB) milling in a $65nm$-thick gold film that had been evaporated beforehand on top of a thin glass substrate. 
Our setup, shown in Fig. 2a consisted of the pulsed laser at $\lambda_0 = 785nm$ (C-Fiber 780 Menlo Femtosecond Erbium Laser, 100mW, 100fs - pulse width) whose beam was expanded to properly fill the aperture of the  microscope objective O1, illuminating the sample. 
The second, oil immersion objective, O2 was brought into a contact with the back (glass) side of our sample in order to produce leakage radiation which was then collected by the tube lens ($100mm$) into the imaging system terminated by a camera (Pixelink, PL-B771U, MONO 27, 1280 x 1024). 
An additional lens was placed on a flipping mount one focal distance ahead of the camera in order to produce the Fourier image of the light distribution. 
Our way to manipulate the incidence angle was by moving an iris placed right in front of the illumination objective in $x$ direction as schematically shown in Figure 2. 
Additionally our setup included the possibility to make a time-resolved imaging of the plasmonic pulses by means of the heterodyne interference method described in the following sections. 
 
\begin{figure}[htbp]
	\centering
		\includegraphics[width=10cm, keepaspectratio] {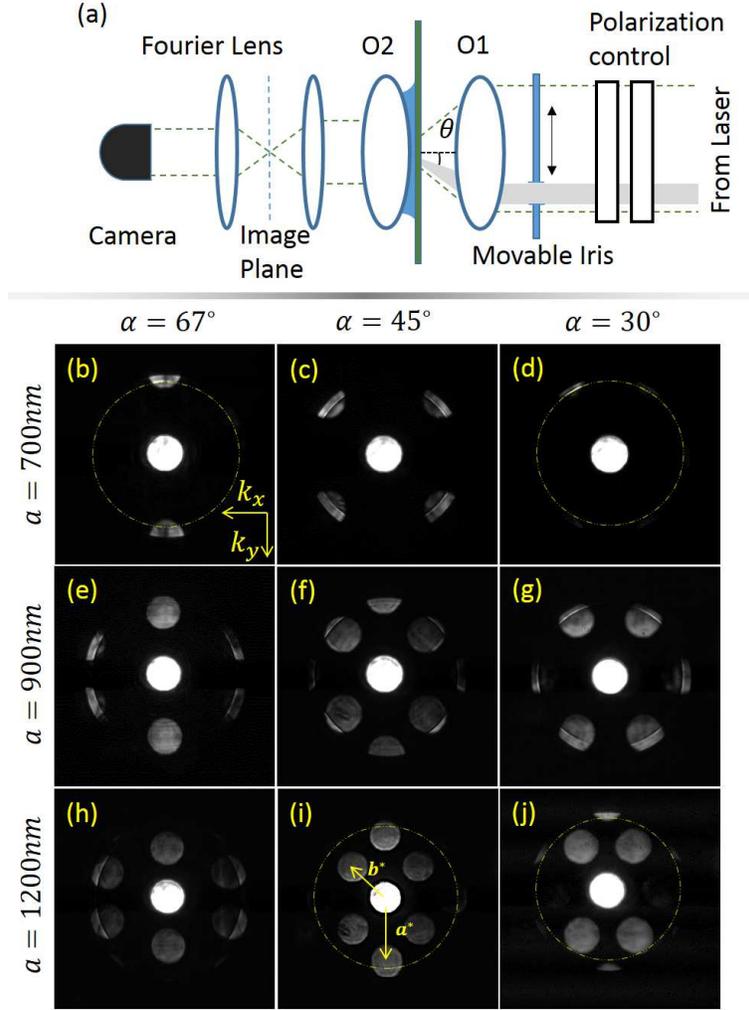}
		\label{Fig2}
		\caption{Inspection of the metasurface $k$-space. (a) The Fourier imaging leakage radiation microscopy setup. The sample is illuminated by an objective with NA = 0.45 while the leakage radiation is extracted by an oil immersion objective O2 with NA = 1.25 coupled with a 100mm tube lens. An additional lens is then used to generate a Fourier image in the camera. A movable slit controls the incidence tilt angle. (b)-(j) Fourier images of metasurfaces with different diagonal angles and different periods as shown. }
	\end{figure}
	
In the first stage we illuminated all the structures at the incidence angle of $\theta = 0^{\circ}$ and with a linear $x$-polarization in order to simply obtain the reciprocal lattice image by using the Fourier lens. 
Below in Fig. 2 the bare images are shown. 
In order to visualize the $k-$space in some pannels we added a dashed yellow circle specifying the plasmonic wavenumber. 
By observing the images one can recognize the diffraction orders (round spots) distributed in the $k$-space according to the reciprocal lattice. 
In Fig. 2i where all the diffraction orders are clearly visible we have schematically marked the reciprocal primitive cell vectors. 
In order to collectively excite a plasmonic wave in a desired direction we aimed to achieve an overlap of the spots representing the vector $\textbf{b}^{\ast}$ with a plasmonic circle. 
With a perfect correspondence to Table 1, in Fig. (b)-(e) the light-SP momentum mismatch is negative which requires a negative tilt of the incident beam. 
On the other hand in Fig. 2(f)-(j) we note that the relevant diffraction orders are inside the plasmonic circle and a positive tilt is needed to compensate for the momentum mismatch. 
Clearly, due to the projection of the incident LS onto the TS of the SPs we expect that the helicity dependence will be the same in the two cases above. 
We wish to test these two cases by choosing one metasurface with $a = 700nm$, $\alpha = 45^{\circ}$ and another structure with $a = 1200nm$, $\alpha = 67^{\circ}$. 

\begin{figure}[htbp]
	\centering
		\includegraphics[width=8cm, keepaspectratio] {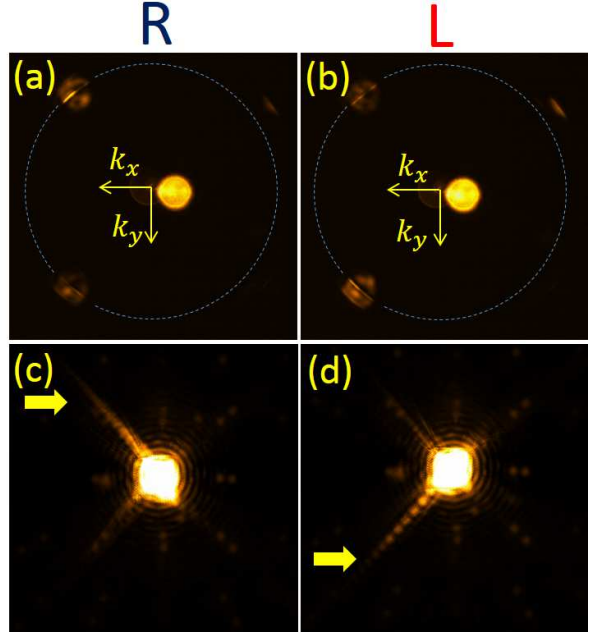}
		\label{Fig3}
		\caption{Spin-selective unidirectional plasmonic excitation with the grating parameters - $a = 700nm$, $\alpha = 45^{\circ}$. (a) and (b) show the k-space for right and left-handed incident polarization respectively and (c) and (d) show the real-space images for these states. The dashes circle guides the eye for the plasmonic wave-vector. Note that the coupling in the k-space is clearly seen as a bright narrow line.}
	\end{figure}
	
Figures 3 and 4 show the measured real and momentum space images of the two structures with the tilt chosen from Table 1 and two circular polarization states. 
The coupling of light to SP wave is clearly seen in the Fourier images as a bright arc inside the diffraction order spot. 
We verify the polarization dependence of this coupling by comparing the LCP and the RCP states. 
Real space images follow the Fourier space and show a unidirectional excitation of the SPs by the metasurface. 
Moreover, by comparing Fig. 3 and 4 one recognizes that in order to excite SP beam in the positive $x$ direction the tilt should be negative in the case of $a = 700nm$ and positive when $a = 1200nm$. 

\begin{figure}[htbp]
	\centering
		\includegraphics[width=8cm, keepaspectratio] {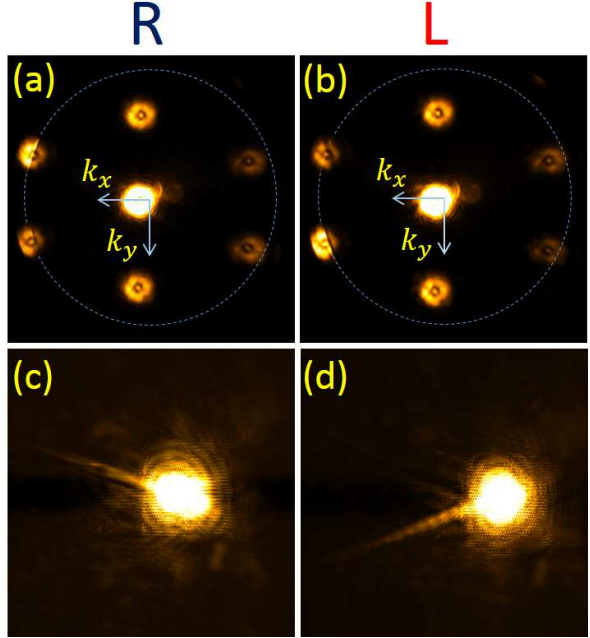}
		\label{Fig4}
		\caption{Spin-selective unidirectional plasmonic excitation with the grating parameters - $a = 1200nm$, $\alpha = 67^{\circ}$. (a) and (b) show the k-space for right and left-handed incident polarization respectively and (c) and (d) show the real-space images for these states.}
	\end{figure}

As expected from the basic theory of the LTS projection the directionality dependence on the incident spin-state stays the same for the two cases. 
The projection of the incident spin onto the transverse spin vector of the SP should vary as $\sim \sin \theta \cos \alpha$, so this is clear that the effect can be more enhanced with higher tilt angles. 
Nevertheless even within the limitations of our simple setup the effect seems to be quite convincing. 

The last experiment is dedicated to the investigation of the dynamics of the excited plasmonic pulse by means of our time-resolved leakage radiation microscopy (LRM)\cite{gorodetski2016}. 
This system incorporates the Mach-Zender type optical interferometry to obtain the spatial-temporal distribution of the SPs (see Fig. 5a). The pulses generated by the laser are then being split into two optical paths. 
The signal path goes through the LRM providing a spatial pulse distribution in the plane of the metal surface. 
The second, reference path is utilized to probe the pulse current position. 
The path can be delayed by a couple of mirrors mounted on a movable stage that is precisely controlled by the computer, therefore the pulse can be captured in different time instants. 
Thus by changing the time-delay we capture the series of interferograms in which the visibility function corresponds to the pulse amplitude envelope. 
More detailed description of the system is available in ref. \cite{gorodetski2016}. 
We use this system to track a $100fs$ plasmonic pulse launched by our metasurface. We choose a grating with $a = 1200nm$ and $\alpha = 67^{\circ}$. 

\begin{figure}[htbp]
	\centering
		\includegraphics[width=8cm, keepaspectratio] {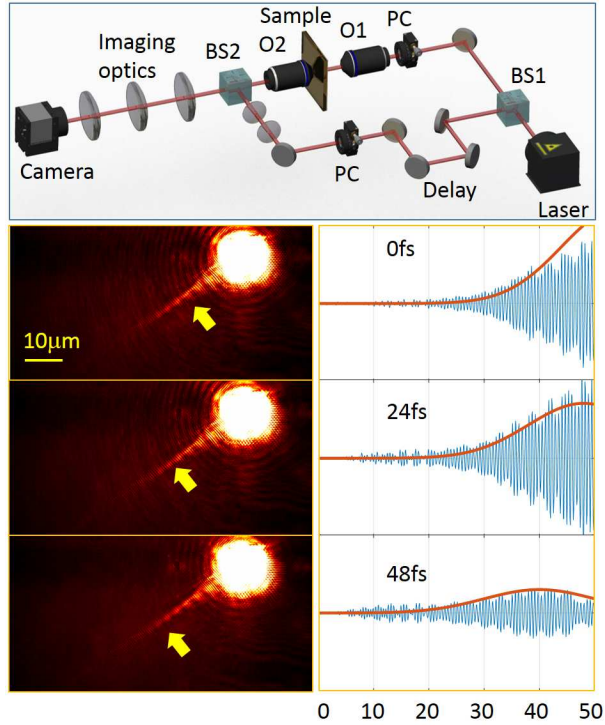}
		\label{Fig5}
		\caption{Time-resolved LRM imaging of the laser pulses launched by the metasurface. (a) The schematics of the setup. 
		The $100fs$ pulse is first split in two parts by the first beam-splitter BS1, after which one part of it arrives to the LRM comprising of the O1  $20X$ objective with $NA = 0.45$ and the oil-immersion O2 $100X$ objective with $NA = 1.25$. 
		The second part of the pulse passes through the delay path and a set of lenses that is used to precisely shape the phase-front of the reference beam to perfectly match with the leakage signal. 
		The polarization control (PC) is used to vary the state of light. (b) A set of three consecutive raw images with a $24fs$ delay in between. 
		(c) The cross-sections of the propagating pulses shown in (b) after image processing and filtering.}
	\end{figure}

In Fig. 5 we present the raw pulse images and the corresponding cross-sections of the propagating pulse after zero-order filtering. The yellow arrow points roughly where the maximum visibility of the fringes is achieved.
The consecutive images corresponding to the time delays of $24fs$ allow us to track the pulse dynamics and extract the velocity characteristics.  
Moreover, by studying the individual frames a phase structure of the pulse excited by our metasurface can be studied. 
By fitting the envelope of the pulse cross-section with a Gaussian shape we achieve the real plasmonic pulse duration $\delta t = 100fs$, and the decay rate of the plasmonic wave from which we derived the imaginary part of the wavenumber $k^{\prime\prime}_{SP} = 0.045$. 
The expected imaginary value of the wavenumber on a flat gold for our wavelength is $0.0185$ but we believe that some roughness of the sputtered surface along with the loss caused by the leakage setup could eventually cause this degradation. 

\section{Summary}

We have proposed a way to collectively excite a unidirectional plasmonic beam with an ability to externally control the propagation direction by means of the incident light polarization. 
We have utilized a metasurface comprising of non-chiral and mirror symmetrical unit cells illuminated by an oblique incidence. 
The proposed scheme has elaborated a combination of a spin-locking effect resulting from the transverse spin of SPs together with a grating momentum matching condition leading to a collective excitation. 
We have tested the metasurface performance in direct and the reciprocal (momentum) space and shown the polarization dependent directionality. 
Moreover, we have demonstrated the temporal dynamics of the waves by means of our time-resolved leakage microscopy by measuring frame-by-frame propagation of a 100fs plasmonic pulse. The presented device is simple in design and fabrication and can undoubtedly be integrated in various photonic circuits and systems. 

\begin{acknowledgement}
This work was supported by the Ministry of Science Technology $\&$ Space, Israel.
\end{acknowledgement}

%
%
%

\providecommand{\latin}[1]{#1}
\makeatletter
\providecommand{\doi}
  {\begingroup\let\do\@makeother\dospecials
  \catcode`\{=1 \catcode`\}=2 \doi@aux}
\providecommand{\doi@aux}[1]{\endgroup\texttt{#1}}
\makeatother
\providecommand*\mcitethebibliography{\thebibliography}
\csname @ifundefined\endcsname{endmcitethebibliography}
  {\let\endmcitethebibliography\endthebibliography}{}

\end{document}